\documentstyle[12pt]{article}
\def\ll {\label}
\def\re{\ref}
\def\c{\cite}

\def\r1{(\ref{$1})}
\def\ot{\otimes}
\def\nn{\nonumber}

\def\th{\theta}
\def\ba{\begin{array}{c}}

\def\ea{\end{array}}

\def\ni{\noindent}
\def\si{\sigma}

\def\De{\Delta}
\def\de{\delta}

\def\ov{\over}
\def\ha{{1\over 2}}

\def\l{\left}
\def\l({\left(}
\def\r){\right)}
\def\r{\right}
\def\rw{\rightarrow}

\def\la{\lambda}
\def\al{\alpha}

\def\be{\begin{equation}}
\def\bc{\begin{center}}
\def\ec{\end{center}}
\def\bit{\begin{itemize}}
\def\eit{\end{itemize}}
\def\ee{\end{equation}}
\def\ed{\end{document}}
\def\bea{\begin{eqnarray}}
\def\eea{\end{eqnarray}}
\begin{document}
\title{ Exact solution of quasi two and three dimensional  
 quantum spin  models
} 

\author{
Anjan Kundu\footnote{E-mail: anjan@tnp.saha.ernet.in} \\  
  Saha Institute of Nuclear Physics,  
 Theory Group \\
 1/AF Bidhan Nagar, Calcutta 700 064, India.
 }
\maketitle
\centerline{cond-mat/9711185 ,\ \ SINP/TNP/97-19} 
\begin{abstract} 

 A class of quasi two and three dimensional  quantum   lattice spin 
models
 with nearest  and next  nearest neighbour 
 interactions is proposed. 
The basic idea of construction is to introduce interactions in
an array of
 $XXZ$ spin chains through twisting transformation. The models 
 belong to   quantum integrable systems allowing 
 explicit 
$R$-matrix solution.   The eigenvalue problem
 can be solved exactly using some symmetry of the
models.

\end{abstract}



Quantum integrable coupled  spin chains with $XXZ$ as well as 
Hubbard interactions   
were introduced 
 through twisting transformation in \c{kunpr96}. Following that exact
eigenvalue solutions of such models and their  extensions
 were obtained in \c{shastry97} and \c{kunhub97}. We generalise here this
concept by coupling  neighbouring   pairs of  
 anisotropic $XXZ$ spin-$\ha$ chains \c{xyz},
 which  results to  a family of   quasi two (2d) and  
 three dimensional (3d) 
  quantum
spin models with   nearest neighbour
(NN)  and next  NN  interactions.
 Since the twisting transformation
preserves  integrability, all the 2d and 3d  systems thus  constructed  
turn out to be quantum integrable.
 Generalising further the useful technique of
  \c{shastry97} for
representing such interactions as the operator dependent  unitary
 transformations, we can solve  exactly  the eigenvalue problem of the
higher dimensional spin models. One should mention here that
   an integrable  2d 
quantum spin model,  its corresponding  classical statistical system
as well as  models with internal degrees of freedom 
were also   obtained earlier following a different rout   
\c{borov}. However  the twisting transformation due to its nice symmetry 
makes our approach much simpler,  allowing easy   construction of quasi
 two, three
and  in principle  any arbitrary dimensional integrable spin models.

Let us start from   spin-$\ha$ operators
$\si^a_{(i,m)}, a=1,2,3;\  i=1,2, \ldots, N;\  m=1,2, \ldots, M;\   $
given at  site $(i,m)$ in a 2d   $N \times M$ lattice
with the commutation relations
\be
[\si_{(j,m)}^a,\si_{(k,n)}^b]= i\epsilon^{abc}
\si_{(j,m)}^c \de_{jk}\de_{nm}
   .\ll{cr}\ee
The simplest 2d quantum  model we propose  may be given by the Hamiltonian
\bea
{\ H} &=& \sum_{j=1}^N \sum_{m=1}^M\si_{(j,m)}^+ \si_{(j+1,m)}^-
\left( \rho_{m0}+i (\rho^{+}_{m1}\si_{(j,m+1)}^3
-\rho^{-}_{m1}\si_{(j+1,m-1)}^3)+\rho_{m2} \  (\si_{(j,m+1)}^3\si_{(j+1,m-1)}^3)
\right) \nonumber \\ &+& 
 \De_m \si_{(j,m)}^3\si_{(j+1,m)}^3
+ h.c.  ,\ll{2dh}\eea
where the parameters $\rho_{ma} $ involve only two independent coupling
constants $\th_{m-1m}$  and $\th_{mm+1}$ between the neighbouring chains 
 as 
\begin{equation}
 \rho_{m0}=\cos 2\th_{m-1m} \cos 2\th_{mm+1}, \ 
 \rho^{\pm}_{m1}=\cos 2 \th_{m\mp 1 m} \sin 2\th_{mm\pm 1}, 
 \rho_{m2}=\sin 2\th_{m-1m} \sin 2\th_{mm+1}. \ll{rho}\end{equation}
 We see
from the above Hamiltonian that it is asymmetric in $x$ and $y$
directions, which  
 gives the model a quasi 2d structure. There are
 $XXZ$ type NN spin interactions along the $x$-direction indicated by the
sites $j$,
while  varieties of different interactions  along the $y$ direction denoted
by the index $m$.

 The terms with coefficients $\rho^{\pm}_{m1} $ represent NN
interaction involving  three spins, while that
 with  $\rho_{m2} $ stands for the next to NN coupling involving four spin
operators. Note that the anisotropy parameters ${\De_m \ov \rho_{m0}}$ 
are different for different chains and moreover depend on
the neighbouring chain couplings $ \th_{mm+1}, \th_{m-1m}$. Therefore 
${\De_m \rw 1}$ limit can not recover the isotropic case. 

We intend to 
 prove that  the model represents   an   exactly  integrable quantum 
system exhibiting a
hierarchy of commuting conserved
 integrals with the Hamiltonian being one of them. For
this we first find the associated quantum $R$-matrix
 satisfying the Yang-Baxter equation (YBE)
\bea  R_{<a><b>}(\lambda -\mu) R_{<a><j>} (\lambda)~
  R_{<b><i>}(\mu) 
=  R_{<b><j>}(\mu)~   R_{<a><j>} (\lambda)~
 R_{<a><b>}(\lambda -\mu),
\nn \\\ll{ybe}\eea
considering  $R^0_{<a><j>}(\la)= \prod_{m=1}^M 
 R^{xxz}_{(a,m)(j,m)}(\la, \eta_m)$ as the $R$-matrix
of $M$ number of integrable 
 noninteracting $XXZ$ spin chains.  $ R^{xxz}_{(a,m)(j,m)}(\la, \eta_m)$
 corresponds  to
the $m$-th chain  acting nontrivially on the space $(I_1\otimes \cdots \ot
I_{m-1} \ot
 V_a \ot \cdots \otimes I_M)\ot(I_1\otimes \cdots \ot I_{m-1} \ot
 V_j \ot \cdots \otimes I_M)$  and given  by its
 well known form \c{xyz}
\begin {equation}
R^{xxz}_{(a,m)(j,m)}(\lambda,\eta_m) = w_0(\la,\eta_m) I_m \ot I_m +
 w_3(\la, \eta_m)
\si^3_{(a,m)} \si_{(j,m)}^3 + w(\eta_m) (\si_{(a,m)}^+ \si_{(j,m)}^- +
 \si^-_{(a,m)} \si^+_{(j,m)}) ,
\ll{Rxxz}\end {equation}
where 
\be
 w_0+w_3={ \sin(\la+\eta_m) }, \    w_0-w_3={ \sin \la }, \   
~~w= \sin \eta_m . 
\ll{trm}\ee
We perform twisting \c{reshit} plus similarity  transformations
 to construct a new $R$-matrix
\be
 R_{<a><j>} (\lambda)=
 F_{<a><j>} G_{<a><j>} R^0_{<a><j>} (\lambda)
 G^{-1}_{<a><j>} F_{<a><j>} 
\ll{r}\end {equation}
where 
\be
 F_{<a><j>}= \prod_m F_{(a,m)(j,{m+1})},\ \  G_{<a><j>}= 
\prod_m   G_{(a,m)(j,{m+1})}
 \ll{fg}\end {equation}
 {with}
 \be F_{(a,m)(j,{m+1})}
=
f^\th_{(a,m)(j,{m+1})}(f^\th_{(a,{m+1})(j,{m})})^{-1} 
\ \ \mbox{and} \  \ \  G_{(a,m)(j,{m+1})}
=
f^\al_{(a,m)(a,{m+1})}f^\al_{(j,{m})(j,{m+1})}  \ll{ffgg}\end {equation}
and the explicit expression through the spin operators  as
\be
f^\th_{(a,m)(j,{m+1})}= \exp [{i \th_{mm+1} \si^3_{(a,m)}\si^3_{(j,m+1)}}]. 
 \ll{f}\end {equation}
and with $ \th_{mm+1}$ replaced by $\al_{mm+1} $
for $f^\al_{(a,m)(j,{m+1})}. $
To show the new $R$-matrix (\re{r}) to be a solution of the YBE (\re{ybe})
  we insert
 it in the equation and notice that due to the specially designed  forms
(\re{ffgg}) and (\re{f}) of  $F$ and $ G,$ the twisting matrices  
can be pushed through all the  $R^0$'s without spoiling their
structures and  canceled from both the sides.
 As a result we are left  
with the  $R^0$ matrices only, which 
satisfy the YBE. For example, for shifting the $F$  factors
through  $R^0$ in the term 
$R^0_{<a><b>}F_{<a><j>}F_{<b><j>}$ appearing in the equation, we notice that
the only term in $R^0$ which does not  commute
 with the $F$ factors is 
  $\prod_{m=1}^M \si_{(a,m)}^+ \si_{(b,m)}^-. $
 However using the
obvious relation
 \be \si_{(a,m)}^\pm \exp
 [ {i \th \hat X(j,m+1) \si^3_{(a,m)}}]= 
\exp [{\pm i \th \hat X(j,m+1)}]
\exp [{i \th \hat X(j,m+1) \si^3_{(a,m)}}]\si_{(a,m)}^\pm \ll{sft}\ee
\ni with arbitrary operator 
 $ \hat X(j,m+1) $, we find  
that 
the extra factor $$\exp [{i(   \th_{m-1m} \si^3_{(j,m-1)} - 
 \th_{mm+1} \si^3_{(j,m+1)}})] $$ produced 
due to the  transition of $F_{(a,m)(j,{m+1})} F_{(a,m-1)(j,{m})}$ is canceled 
exactly by  the  factor
created due to  
$F_{(b,m)(j,{m+1})}F_{(b,m-1)(j,{m})}.$
 Thus $R^0_{<a><b>}$ remains unchanged after
taking  the $F$ factors through it.
 Similar reasoning holds for $R^0_{<a><j>}$
and $ 
R^0_{<b><j>}$. The factors related to the 
 similarity transformation:   $
G_{(a,m)(b,{m+1})}^{-1}G_{(a,m)(j,{m+1})}$ etc.  on the other hand,
 are partially canceled among themselves and
the remaining ones commute trivially with the $R^0$ matrices. 
This shows that the
transformed $R$-matrix (\re{r}) is also a solution of the YBE, which  
 proves  
the quantum integrability of the system and  
 guarantees that the  transfer matrix   
$~~\tau(\la)=tr_{<a>} \left(\prod_{j=1}^N R_{<a><j>}(\la) \right)~
$  would generate  mutually commuting set of conserved operators $C_n=
({\partial^n \ov \partial \la^n } \log \tau (\la))_{\mid \la=0}$ 
\c{baxter,xyz}.

 We  construct  now the Hamiltonian in the   explicit form 
supposing  $H \equiv C_1=
  \tau' (0)\tau^{-1} (0)$
and using the 
definition of $\tau(\la)$ along with the expressions (\re{r}) and
 (\re{Rxxz}). 
 Notice  an  important property of the $R$-matrix  (\re{r}): 
\bea
 R_{<a><j>} (0)&=&c
 F_{<a><j>} G_{<a><j>} P_{<a><j>} 
 G^{-1}_{<a><j>} F_{<a><j>} \nn \\ &=& c
 F_{<a><j>} G_{<a><j>} 
 G^{-1}_{<j><a>} F_{<j><a>} P_{<a><j>}=c P_{<a><j>},
\ll{rP}\eea
which  follows easily from that  of $
R^0_{<a><j>}(0)= \prod_{m=1}^M 
 \sin \eta_m P_{(a,m)(j,m)}= c
  P_{<a><j>} $ and the symmetry   $  
 F_{<j><a>}= F_{<a><j>}^{-1}, \ \  G_{<j><a>}= G_{<a><j>}.$
Using this along with the property of the permutation operator 
like
$P_{<a><j>} R'_{<a><j+1>}(0)= R'_{<j><j+1>}(0)P_{<a><j>},$ \ \ 
and $P_{<a><j>}^2=I $
we get
\be
\tau (0)
 =c \left( (\prod_{m=1}^MP_{(j,m)(j+1,{m})})
  P_{<j><j+2>}\ldots\right)tr_a(P_{<a><j>}) 
\ll{tau0}\ee
and
\bea
\tau' (0)
&=&c \sum_{j=1}^N \sum_{m=1}^M 
 (\prod^m_{l=m-1}(F_{(j,l)(j+1,{l+1})} G_{(j,l)(j+1,{l+1})})
R'^{xxz}_{(j,m)(j+1,{m})}(0)
\prod_{n\not = m}P_{(j,n)(j+1,n)}
 \nn \\ & & \prod^m_{l=m-1}F_{(j,l)(j+1,{l+1})} G^{-1}_{(j,l)(j+1,{l+1})})
   P_{<j><j+2>}\ldots 
\times tr_a(P_{<a><j>}) ,
\ll{tau1}\eea
yielding 
\be
H
= \sum_{j=1}^N \sum_{m=1}^M S^m_{jj+1}
R'^{xxz}_{(j,m)(j+1,{m})}(0)P_{(j,m)(j+1,{m})}(S^m_{jj+1})^{-1}, \ \ S^m_{jj+1}=
\prod^m_{l=m-1}F_{(j,l)(j+1,{l+1})} G_{(j,l)(j+1,{l+1})}.
\ll{h}\ee
Expressions (\re{Rxxz}),(\re{f}) through spin operators and their properties 
(\re{sft}) 
 reduce  (\re{h}) finally to 
a family of 2d quantum spin models
\bea
{\ H} &=& \sum_{j=1}^N \sum_{m=1}^M\si_{(j,m)}^+ \si_{(j+1,m)}^-
 \exp (i[ (\th_{mm+1}+\al_{mm+1})\si_{(j,m+1)}^3
- (\th_{m-1m}+\al_{m-1m})\si_{(j+1,m-1)}^3  \nonumber \\&+&
 (\th_{mm+1}-\al_{mm+1}) 
 \si_{(j+1,m+1)}^3 - (\th_{m-1m}-\al_{m-1m})    \si_{(j,m-1)}^3])
 +\De_m \si_{(j,m)}^3\si_{(j+1,m)}^3  + h.c., \nn \\
  \ll{2dgh}\eea
 parametrised by the set $\{\th_{mm+1}\},
\{\al_{mm+1}\}$.
This also   establishes that  the family of models   (\re{2dgh}) belongs  to 
 the hierarchy of 
quantum integrable systems. Choosing the coupling parameters $\th, \al$
differently we can generate quasi 2d integrable models with rich varieties
of interchain interactions. The simplest choice $
\al_{mm+1}=\th_{mm+1}
,$ 
using  $(\si^3_{(a,m)})^2=I$  clearly yields the Hamiltonian (\re{2dh}), while 
$ 
\al_{ll+1}=(-1)^l\th_{ll+1}$ leads to the models like
\bea
{\ H} = \sum_{j=1}^N \sum_{m=1}^M(\si_{(j,m)}^+ \si_{(j+1,m)}^-
S_m
 +\De_m \si_{(j,m)}^3\si_{(j+1,m)}^3)+ h.c.
  ,\ll{2dgh1}\eea
with $S_m  $ taking  alternate expressions  as
\bea 
S_m &=& \exp (2i[ \th_{mm+1}\si_{(j,m+1)}^3
 - \th_{m-1m}   \si_{(j,m-1)}^3])
,\nn \\
S_{m+1} &=& \exp (2i[ \th_{m+1m+2}\si_{(j+1,m+2)}^3
 - \th_{mm+1}   \si_{(j+1,m)}^3]).
\ll{s}\eea
One may also choose further the $M$ number of parameters $\th_{mm+1} $ 
differently to get other types of interacting 2d models.

Next we address the eigenvalue problem:
$
H \mid \Psi>= E \mid \Psi> $ and show that by exploiting the 
 symmetry of the operators  $F_{(j,m)(j+1,{m+1})},  G_{(j,m)(j+1,{m+1})}$
 appearing in 
(\re{h})   and  using the known  
 results for the 
 $XXZ$ spin chains one can solve this problem exactly. 
Indeed, using  the  relations like (\re{sft})
one can conclude interestingly that  our Hamiltonian  (\re{h})
can be reduced  to the form
$H=S H_0 S^{-1},$ where   
$H_0$ is a collection of exactly solvable $XXZ$ chains and  
$S$ is an unitary operator given by
\bea
S=\prod_{{k,m,i,j} (i<j)}  f^\al_{(k,m)(k,{m+1})} 
f^\th_{(i,m)(j,{m+1})}(f^\th_{(i,{m+1})(j,{m})})^{-1}
  ,\ll{ss}\eea
with $f^\th,f^\al$ as in (\re{f}). Note that for the particular 
$m=2$ case and $\al=0$ the result is consistent
with \c{shastry97}. 
Therefore  the eigenvalue problem of the quasi 2d  Hamiltonian  (\re{2dh})
can be reduced to that of $H_0$ with the same energy spectrum
$ E=\sum_m^M \sum_k^{M_m} \cos p_{k_m}$, where  $M_m$ are the number of
spin excitations in the $m$-th chain. The eigenfunction on the other hand
may be given by $ S (\prod_m \mid \Psi^{(m)}_0> ),
 $ where 
$ \mid \Psi^{(m)}_0>  $  is the known 
solution for a single $XXZ$ chain
solvable  by the Bethe ansatz \c{xxz}.
Note that, though our model can be  solved  through  integrable  
$XXZ$ chains,  the results  are not really  equivalent. 
 Indeed,
the energy spectrum 
depends on  the values of the unknown momentum parameters 
$p_{k_m}$, which in turn
can be  determined by the Bethe equations and  
these equations are not
the
same for the two models. In the 2d model due to the operator $S$ extra
factors like 
$\  \exp [i ( \th ^{mm+1}M_{m+1}- \th ^{m-1m}M_{m-1})] \ $ appear in the
determining Bethe equations
signifying interactions between the neighbouring chains and this 
 gives the set of  momentum parameters a different
value. The nature of the eigenfunction is also changed due to the same
reason. At different sectors in the configuration space 
different phase factors arise in the wave functions due to the
operator nature of $S$ and at the boundaries of the sectors the 
wave functions suffer phase jumps resulting the occurrence of 
 discontinuity at 
coinciding points. This unusual feature was  observed also 
in case of  the twisted Hubbard model \c{kunhub97}.  

We extend now this idea to construct quasi 3d quantum spin
models by defining the $R$-matrix (\re{r}) with transforming operators 
\be
 F_{<a><j>}= \prod_{m,p} F^{(m)}_{(a,m,p)(j,{m+1},p)}
 F^{(p)}_{(a,m,p)(j,{m},p+1)},\ \  G_{<a><j>}= 
\prod_{m,p}   G^{(m)}_{(a,m,p)(j,{m+1},p)}
 G^{(p)}_{(a,m,p)(j,{m},p+1)}
 \ll{fg3}\end {equation}
Here as before the index $j=1, \ldots, N$  denotes the site number in the
$x$-direction, $m=1, \ldots, M$ stands for the chain number along the
$y$-direction, while the additional index   
$p=1, \ldots, L$
indicates the layer number along the $z$-direction. 
Therefore we may extend our above definitions to the 3d case as
 $$ F^{(p)}_{(a,m,p)(j,{m},p+1)}
=
f^{\th_{pp+1}}_{(a,m,p)(j,{m},p+1)}(f^{\th_{pp+1}}_{(a,m,{p+1})(j,{m},p)})^{-1} 
$$ {and} \be  G^{(p)}_{(a,m,p)(j,{m},p+1)}
=
f^{\al_{pp+1}}_{(a,m,p)(a,m,{p+1})}
f^{\al_{pp+1}}_{(j,m,p)(j,m,{p+1})}
 \ll{ffgg3}\end {equation}
with 
\be
f^{\th_{pp+1}}_{(a,m,p)(j,m,{p+1})}= \exp [{i
 \th_{pp+1} \si^3_{(a,m,p)}\si^3_{(j,m,p+1)}}]. 
 \ll{f3}\end {equation}
for fixed chain index $m$ and $\th_{pp+1} $
as the interlayer coupling. Analogous 
 relations as (\re{ffgg3})
hold also for $F^{(m)}_{(a,m,p)(j,{m+1},p)}$ with fixed layer index $p$.
All the above arguments  for proving the $R$-matrix as
the
YBE solution also  go through for  the present extension 
 and one can
generate the related Hamiltonian of a family of 
 quantum integrable 3d spin models.
In the simplest case of $\al_{ab}=\th_{ab} $ the explicit form of such models 
may be given as
\bea
{\ H} &=& \sum_{j=1}^N \sum_{m=1}^M \sum_{p=1}^L
(\si_{(j,m,p)}^+ \si_{(j+1,m,p)}^-
S_{m,p}^{jj+1} 
 +\De_m \si_{(j,m,p)}^3\si_{(j+1,m,p)}^3)+ h.c.
  \ll{2dh3}\eea
with the interchain and interlayer interactions given as
\be
S_{m,p}^{jj+1}
=\exp (2i[ (\th_{mm+1}\si_{(j,m+1,p)}^3
- \th_{m-1m}\si_{(j+1,m-1,p)}^3)+
(\th_{pp+1}\si_{(j,m,p+1)}^3
- \th_{p-1p}\si_{(j+1,m,p-1)}^3)]).
\ll{s3}\ee
Note that due to the factorised form of
$$
S_{m,p}^{jj+1}=S_{(m+1,p)}^{j}( S_{(m-1,p)}^{j+1})^{-1}
 S_{(m,p+1)}^{j}( S_{(m,p-1)}^{j+1})^{-1}$$
with \be
 S_{(m,p+1)}^{j}
=\exp (2i[
\th_{pp+1}\si_{(j,m,p+1)}^3])=\cos 2\th_{pp+1} +i
 \si_{(j,m,p+1)}^3 \sin 2\th_{pp+1}
\ll{sp3}\ee
etc. such interactions take place between NN and next  NN chains in the
same plane (with fixed $p$) as well as between NN and next NN layers in the
same chain (with $m$ fixed) and finally between such terms themselves.
 
The eigenvalue problem of such 3d models can also be solved similar to 
 the 2d case 
by transforming the Hamiltonian to the noninteracting array of $XXZ$
spin chains through unitary transformation.

Thus we have constructed  and solved   a class of  
quasi 2d  and 3d quantum spin models and shown also their exact integrability.
It should be  obvious from our construction   
that the approach allows further extension of
 such integrable models to any arbitrary
dimension. 
Unlike the long ranged systems
 \c{frahm},
 the present class of 2d  models includes only
nearest and next nearest neighbour interactions. Therefore 
 such models, possibly with more interesting twisting operators,
 might be helpful in
  constructing integrable  
 higher dimensional   
 physically significant  models, which could not be achieved through the
approach of tetrahedron equation \c{tetra}.  
 It would  also be   important  to construct   
higher dimensional Hubbard like
 models through  similar technique \c{kun98}.

\ni

\end{document}